\documentclass[a4paper,twocolumn]{article}

\usepackage{amsmath}
\usepackage{graphicx}
\usepackage{cite}
\usepackage{hyperref}
\usepackage{epstopdf}

\begin{document}

\title{An analytical prediction of the bifurcation scheme of a clarinet-like
instrument: Effects of resonator losses}
\author{P.-A. Taillard \\
%EndAName
University of Applied Sciences and Arts Northwestern Switzerland \\
Musik-Akademie Basel. Academy of Music.\\
Schola Cantorum Basiliensis, Leonhardsstr. 6 CH-4051 Basel\\
J. Kergomard \thanks{%
Tel 33 491164381, Fax 33 491228248, kergomard@lma.cnrs-mrs.fr} \\
LMA, CNRS, UPR 7051, Aix-Marseille Univ, Centrale Marseille,\\
F-13402 Marseille Cedex 20, France}
\maketitle

\begin{abstract}
The understanding of the relationship between excitation parameters and
oscillation regimes is a classical topic concerning bowed string
instruments. The paper aims to study the case of reed woodwinds and attempts
to find consequences on the ease of playing.

In the minimum model of clarinet-like instruments, three parameters are
considered: i) the mouth pressure, ii) the reed opening at rest, iii) the
length of the resonator \ assumed to be cylindrical. Recently a
supplementary parameter was added: the loss parameter of the resonator
(using the \textquotedblleft Raman model\textquotedblright , that considers
resonator losses to be independent of frequency). This allowed explaining
the extinction of sound when the mouth pressure becomes very large. The
present paper presents an extension of the paper by Dalmont et al (JASA,
2005), searching for a diagram of oscillation regimes with respect to the
reed opening and the loss parameter. An alternative method is used, which
allows easier generalization and simplifies the calculation. The emphasis is done on the emergence
bifurcation: for very strong losses, it can be inverse, similarly to the
extinction one for weak losses. The main part of the calculations are
analytical, giving clear dependence of the parameters. An attempt to deduce
musical consequences for the player is given.
\end{abstract}

%\begin{doublespace}


\section{Introduction}

The understanding of the relationship between excitation parameters and
oscillation regimes is a classical topic concerning bowed string
instruments: for instance Shelleng \cite{shelleng}, or Guettler \cite%
{guettler}, or Demoucron et al \cite{causse} proposed 2D diagrams with
respect to either bow force and bow position, or bow position and bow
velocity.\ For reed instruments, this kind of diagrams are less numerous: in
Ref. \cite{Dalmont:05}, Dalmont et al proposed a diagram with respect to
excitation pressure and reed opening, and recently Almeida et al \cite%
{almeida} proposed a diagram with respect to blowing pressure and lip force,
related to the reed opening.

In the minimum model of reed, clarinet-like instruments, three parameters
are considered: i) the mouth pressure, ii) the reed opening at rest, iii)
the length of the resonator \ assumed to be cylindrical. In Ref. \cite%
{Dalmont:05}, a supplementary parameter was added: the loss parameter of the
resonator (using the \textquotedblleft Raman model\textquotedblright , that
considers resonator losses independent of frequency). This allowed
explaining the extinction of sound when the mouth pressure becomes very
large. We notice that the agreement of the theoretical results \ with
experimental results was satisfactory (see Ref. \ \cite{Dalmont:07}).

The objective of the present paper is to revisit the paper by Dalmont et al%
\cite{Dalmont:05}: the focus is the search for a diagram of oscillation
regimes of reed instruments with respect to two parameters: the reed
opening, and the loss parameter. The choice of these two parameters is
justified by the fact that the third parameter, the blowing pressure, is the
easiest to modify for the instrumentalist. The elements of this diagram were
rather complete in Ref. \cite{Dalmont:05}, but phenomena occurring for
strong losses, especially at the emergence of the sound, were not
investigated.

The use of simplified models for the prediction of the oscillation regimes
is classical for musical instruments producing self-sustained oscillations.
For the calculation of the instability thresholds, linearization was used
(see Wilson and Beavers \cite{Wilson} or Silva et al \cite{silva}), while
for ab initio computation, the iterated map scheme was studied (see Mc
Intyre et al \cite{McIntyre:83}, Maganza et al \cite{Maganza:86}, Taillard
et al \cite{taillard}). The interest of the model chosen in the present
paper is that analytical formulas are possible, given clear dependence of
the parameters (for other models, numerical calculations could be possible
by using similar basic ideas, using for instance continuation methods\cite%
{karkar} or time-domain methods \cite{guillem, tachi}).

The analytical calculations presented hereafter are slightly different from
those of Ref. \cite{Dalmont:05}. They also are limited to the limit cycles
corresponding to the two-state oscillating regime, but are based upon a
generalization of the fact that for this regime, when no losses are present,
the flow rate is a constant. This regime is the most similar to what
musicians consider as a \textquotedblleft normal" sound. In particular the
method allows studying the character of the emergence and extinction
bifurcations of this regime, which are important properties related to the
possibility to play pianissimo or not, and more generally to the ease of
playing. In Ref. \cite{Dalmont:05} it was shown that the extinction
bifurcation can be direct or inverse (supercritical and subcritical, respectively, see Refs; \cite{115,171,252}); here it is shown that this is true
also for the emergence bifurcation.

In section \ref{model}, the basic model of Ref. \cite{Dalmont:05} is
reminded (see also \cite{Kergomard:95}), and a treatment of the problem
based upon a unique quantity, the pressure difference $\Delta p,$ is
presented in section \ref{recu}.\ This leads to a simple graphical analysis
of the two-state regime, explained in section \ref{exis}, yielding a proof
that it cannot exist with reverse flow, and an easy calculation method. Then
some blowing pressure thresholds (stability, existence, ...) are calculated
with respect to the parameters of interest (loss, reed opening).

In section \ref{CT} the thresholds related to the instability of the regimes
are calculated. Then by making two thresholds equal, the mouth pressure can
be eliminated and limits of existence and stability of the static and
two-state regimes are found in Section \ref{DIA}: this allows drawing the
diagram sought. Finally a discussion is proposed concerning the existence of
oscillating regimes (Section \ref{disc}), with an attempt to consider more
realistic models and a discussion about musical consequences.

\section{The model and its parameters\label{model}}

We briefly remind the basic elements of the model, the non-linear
characteristic of the exciter, and the origin of the iteration method,
thanks to a simplified treatment of the resonator.

\subsection{\protect\bigskip Nonlinear characteristics of the entering flow}

In a quasi static regime, the flow $U$ entering the resonant cavity is
modeled with the help of an approximation of the Bernoulli equation, as
discussed e.g. in \cite{Hirschberg:90}. Comparison with experiment can be
found in Ref. \cite{Dalmont:03}. We note $P_{int}$ the acoustic pressure
inside the mouthpiece, assumed to be equal to the one at the output of the
reed channel, $P_{m}$ the pressure inside the mouth of the player. For small
values of the difference:
\begin{equation}
\Delta P=P_{m}-P_{int}\text{ ,}  \label{delta-P}
\end{equation}%
the reed remains close to its equilibrium position, and the flow $U$ is
proportional to $sign(\Delta P)\sqrt{\left\vert \Delta P\right\vert }$; for
larger values of this difference, the reed moves and, when the difference
reaches the closure pressure $P_{c}$, it completely blocks the flow (the reed is beating).\ These
two effects are included by assuming that if $\Delta P\leq P_{c}$ the flow $%
U $ is proportional to $sign(\Delta P)\sqrt{\left\vert \Delta P\right\vert }%
\left[ P_{c}-\Delta P\right] $, and if $\Delta P>P_{c}$, the flow vanishes.\
Introducing the dimensionless quantities:
\begin{equation}
\begin{array}{l}
p=P_{int}/P_{c}\text{ \ \ ; \ }u=UZ_{c}/P_{c} \\
\gamma =P_{m}/P_{c}\text{ ; \ }\gamma _{c}=P_{c}/P_{c}=1\text{.}%
\end{array}
\label{defs}
\end{equation}%
where $Z_{c}=\rho c/S$ is the characteristic acoustic impedance of the
cylindrical resonator, having the cross section $S$ ($\rho $ is the density
of air, $c$ the velocity of sound), we obtain:
\begin{equation}
u=\zeta f(\Delta p)\text{ }  \label{e1}
\end{equation}%
with%
\begin{eqnarray}
\Delta p &=&\Delta P/P_{c}=\gamma -p\text{ ;}  \label{2b} \\
f(\Delta p) &=&0\text{ \ \ if \ \ \ \ \ }\Delta p>1\text{\ ;}  \label{e2} \\
f(\Delta p) &=&sign(\Delta p)\;(1-\Delta p)\sqrt{\left\vert \Delta
p\right\vert }\text{ \ \ if }\Delta p<1\text{.}  \label{e2a}
\end{eqnarray}%
The parameter $\zeta $ characterizes the intensity of the flow and is
defined as:
\begin{equation}
\zeta =\frac{cS_{op}}{S}\sqrt{\frac{2\rho }{P_{c}}}\text{ ,}  \label{zeta}
\end{equation}%
where $S_{op}$ is the opening cross section of the reed channel at rest.\ $%
\zeta $ is inversely proportional to square root of the reed stiffness,
contained in $P_{c}$. In real clarinet-like instruments, typical values of
the parameters are $\gamma \in \left[ 0,1.5\right] $; $\zeta \in \left[
0.1,0.5\right] $ ; values $\zeta >1$ will not be considered here, since they
correspond to multi-valued functions to be solved (see Ref. \cite{taillard}%
), and this case does not seem very realistic in practice for clarinet-like
instruments. The function $f(\Delta p)$ is obviously non-analytic; it is
made of three separate analytic pieces, with a singular point at $\Delta p=0$%
, and its derivative is discontinuous at $\Delta p=1$.

\subsection{Resonator model}

The resonator of length $\ell $ is assumed to be cylindrical, with zero
terminal impedance. Using the d'Alembert decomposition, a change in
variables at the entry of the resonator can be chosen as:%
\begin{equation}
p(t)=p^{+}(t)+p^{-}(t)\text{ \ ; }u(t)=p^{+}(t)-p^{-}(t)  \label{e3a}
\end{equation}%
with the following relationship between incoming wave $p^{-}(t)$ and
outcoming wave $p^{+}(t)$:%
\begin{equation}
p^{-}(t)=-\lambda p^{+}(t-2\ell /c)  \label{e3}
\end{equation}%
where $\lambda $ is the loss parameter, assumed to be independent of
frequency. This is a strong assumption, necessary to obtain square signals:
for certain initial conditions, all quantities remain constant in the time
interval $2nl/c<t<2(n+1)l/c$. The approximation is rough for certain
characteristics of the signal, such as the spectrum, but it is useful for
the study of the existence, stability and amplitude of the produced sound.
With this assumption, the resonator is characterized by a unique
(recurrence) relation:%
\begin{equation}
p_{n}^{-}=-\lambda p_{n-1}^{+}  \label{e4}
\end{equation}%
where $2\ell /c$ is the time unit. As discussed further in section \ref{RS},
losses can occur either at the extremity (radiation) or during propagation
(e.g. in the boundary layers): in the latter case, if $\alpha $ is the
attenuation constant per unit length, $\lambda =\exp (-2\alpha \ell ),$ and
the dimensionless input impedance at zero frequency is
\begin{equation}
\mu \overset{def}{=}\tanh (\alpha \ell )=\frac{1-\lambda }{1+\lambda }\text{
,}\text{ \ or }\lambda =\frac{1-\mu }{1+\mu }  \label{e49}
\end{equation}%
while the input impedance at the operating frequency is $1/\mu $. In what
follows, the losses are characterized by the parameter $\mu $, varying
between $0$ (no losses) $\ $and $1$ (very strong losses, no wave
reflection). This parameter, the reed opening $\zeta $ and the mouth
pressure $\gamma $ are the three parameters of the problem. Several
combination parameters will be useful:%
\begin{equation}
\beta \overset{def}{=}\zeta \mu ;\text{ \ }\beta _{1}\overset{def}{=}\zeta
^{-1}\mu \text{ \ or }\zeta ^{2}=\beta /\beta _{1}\text{ \ ; }\mu ^{2}=\beta
\beta _{1}\text{ .}  \label{e63}
\end{equation}%
$\beta $ is proportional to the input impedance at zero frequency, while $%
\beta _{1}$ is proportional to the input admittance at the operating
frequency. Two pairs of parameters can be used: either $(\zeta ,\mu )$ or $%
(\beta ,\beta _{1}).$ Notice that because $\zeta $ and $\mu $ are smaller
than unity, $\beta <1$ and $\beta <\beta _{1}$, and $\beta \beta _{1}=\mu
^{2}<1$. Other parameters will be useful\footnote{%
In Ref. \cite{Dalmont:05}, the parameters $\gamma $ and $\zeta $ are with
dimension, except in the appendix, and are denoted $p_{m}$ and $u_{A}$,
respectively. $\beta $, $\beta _{1}$ and $\beta _{2}$ are defined in the
same way than in the present paper.}:%
\begin{equation}
\beta _{2}\overset{def}{=}\frac{2\beta _{1}}{1+\beta \beta _{1}}=\frac{\tanh
2\alpha \ell }{\zeta }\text{ \ ; }\beta _{3}\overset{def}{=}\frac{1}{2\beta }%
-\frac{3\beta _{1}}{2}.  \label{e65a}
\end{equation}

\section{Equations for transients and limit cycles\label{recu}}

\subsection{Recurrence for the pressure difference}

\bigskip Using Eqs. (\ref{e1}), (\ref{e3a}) and (\ref{e49}), the recurrence
relation (\ref{e4}) can be rewritten with respect to the quantities $\Delta
p_{n}\overset{def}{=}\gamma -p_{n}$ and $u_{n}$. The result is:%
\begin{equation}
2\gamma =(1+\mu )(\Delta p_{n}+u_{n})+(1-\mu )(\Delta p_{n-1}-u_{n-1}).
\label{PA1}
\end{equation}%
Because the flow rate $u=\zeta f(\Delta p)$ is a function of the pressure
difference $\Delta p,$ this relation is a recurrence for the quantity $%
\Delta p$, equivalent to the recurrence used in Ref. \cite{taillard} for the
quantity $p^{+}$:%
\begin{eqnarray}
\Delta p_{n} &=&H^{-1}\left[ K(\Delta p_{n-1})\right] \text{ \ \ with}
\label{PA2} \\
H(x) &=&x+\zeta f(x)\text{;\ \ }  \notag \\
K(x) &=&\gamma (1+\lambda )-\lambda (x-\zeta f(x)).
\end{eqnarray}
The inverse of function $H$ can be found in Ref. \cite{taillard} (Appendix
A).

\subsection{\protect\bigskip Basic equations for the static regime\label%
{besr}}

Eq. (\ref{PA1}) is interesting in particular for the calculation of the
limit cycles. For the static regime, $\Delta p$ is a constant, then%
\begin{equation}
\gamma =\Delta p+\beta f(\Delta p)\overset{def}{=}h(\Delta p).  \label{e8}
\end{equation}%
It is possible to calculate $\Delta p$ from the value of $\gamma $, or
vice-versa. Concerning the stability, if the iteration function is denoted $%
g(x)=H^{-1}\left[ K(x)\right] ,$ the classical stability condition is $%
\left\vert g^{\prime }(\Delta p)\right\vert <1$. Because $K(x)=H\left[ g(x)%
\right] ,$ $dK/dx=(dH/dg)(dg/dx)$, and the condition can be written as%
\footnote{%
Notice that $\left\vert A/B\right\vert ^{2}<1$ is equivalent to $%
(A-B)(A+B)<0.$}:%
\begin{eqnarray}
\left[ \lambda \frac{1-\zeta f^{\prime }(\Delta p)}{1+\zeta f^{\prime
}(\Delta p)}\right] ^{2} &<&1\text{ \ }  \label{e10} \\
\text{\ or \ }\left[ f^{\prime }(\Delta p)+\beta _{1}\right] \left[ 1+\beta
f^{\prime }(\Delta p)\right]  &>&0\text{\ }
\end{eqnarray}%
(see Eq. (\ref{e63})).

\subsection{Basic equations for the two-state regime}

For the two-state regime, because $\Delta p_{n+1}=\Delta p_{n-1}$, the
following expression is found by eliminating $\gamma $ from the equation (%
\ref{PA1}) written for the pairs ($n+1$, $n$) and ($n$, $n-1$):%
\begin{eqnarray}
h_{1}(\Delta p_{n}) &=&h_{1}(\Delta p_{n-1})\text{ \ \ with}  \label{e889} \\
&&h_{1}(X)\overset{def}{=}\beta _{1}X+f(X).
\end{eqnarray}%
An important property of the two-state regime is the square shape of the
signal, which can be decomposed into the sum of a mean value $p_{mean}$
(zero frequency component) and an acoustic component $p_{ac}$ (sum of the
odd harmonics of the operating frequency), with zero mean value:%
\begin{eqnarray}
p_{n} &=&p_{mean}+p_{ac}\text{ with }p_{mean}=\frac{1}{2}(p_{n}+p_{n-1})%
\text{ }  \notag \\
&&\text{and }p_{ac,n}=\frac{1}{2}(p_{n}-p_{n-1}).  \label{e887}
\end{eqnarray}%
Notice that $p_{ac,n}=-p_{ac,n-1}$, and that a similar equation can be
written for the flow rate. Eq. (\ref{e889}) generalizes the result obtained
when losses are ignored ($\mu =0)$, i.e. the constant flow rate, and it is
nothing else than the input impedance relation for the acoustic component:%
\begin{equation}
\mu (p_{n}-p_{n-1})=u_{n}-u_{n-1}.  \label{e886}
\end{equation}

It is possible to calculate the values of the two states without knowledge
of $\gamma $, starting e.g. with the value of $\Delta p_{n}$: $\Delta
p_{n\pm 1}$ and $\gamma $ are successively deduced from Eqs. (\ref{e889})
and Eq. (\ref{PA1}). Adding the two equations (\ref{PA1}) for the pairs ($%
n+1 $, $n$) and ($n$, $n-1$), it is obtained:
\begin{eqnarray}
\gamma _{ij} &=&\frac{1}{2}\left[ h(\Delta p_{i})+h(\Delta p_{j})\right]
\text{, or}  \label{e921a} \\
\gamma _{ij} &=&\frac{1}{2}(\Delta p_{i}+\Delta p_{j})(1-\beta \beta
_{1})+\beta h_{1}\text{, }  \label{e921}
\end{eqnarray}%
with $h_{1}=h_{1}(\Delta p_{i})=h_{1}(\Delta p_{j})$ and $i=n$, and $j=n\pm
1 $. Similarly for regimes with more than two states, it could be possible
to start the calculation from a given state, and to deduce the other states,
the prior knowledge of $\gamma $ being unnecessary. Eq. (\ref{e921a}) is the
input impedance relation for the mean value component: $p_{n}+p_{n-1}=\mu
(u_{n}+u_{n-1}).$

For the two-state regime, the stability condition is $\left\vert g^{\prime
}(\Delta p_{i})g^{\prime }(\Delta p_{j})\right\vert <1$, and after some
algebra, the result of Ref. \cite{Dalmont:05} is found (see Eq. (\ref{e65a})):%
\begin{eqnarray}
C &<&\beta _{2}\text{ or \ }C>\text{ }\frac{1}{\zeta ^{2}\beta _{2}},\text{ }
\label{e301} \\
\text{with }C &=&-\frac{f^{\prime }(\Delta p_{i})+f^{\prime }(\Delta p_{j})}{%
1+\zeta ^{2}f^{\prime }(\Delta p_{i})f^{\prime }(\Delta p_{j})}.
\end{eqnarray}

\section{Existence of the static and two-state regimes\label{exis}}

The previous results can be applied whatever the shape of the function $%
f(\Delta p).$ The present section investigates the existence of the static
and two-state regimes for the particular shape of the function given by Eqs.
(\ref{e2}) and (\ref{e2a}).

\subsection{Static regime}

For negative $\Delta p$, the function $f(\Delta p)$ is negative too,
therefore the static regime does not exist for negative flow (and positive
excitation pressure $\gamma $). For $\Delta p>1$, the static regime exists
for $\gamma =\Delta p>\gamma _{c}=1$. The pressure $p=\gamma -\Delta p$
vanishes: this is obvious because the reed closes the input of the resonator.

Otherwise, for a non-beating reed, the study of the function in the
right-hand side of Eq. (\ref{e8}) shows that it increases from $0$ to $1$
when $\Delta p$ increases from $0$ to $1.$ Therefore a unique solution%
\footnote{%
The expression for the static pressure in the mouthpiece is the following:%
\begin{eqnarray*}
p_{s} &=&\gamma -\frac{1}{9}\left[ \frac{1}{\beta }+2\sqrt{3+\frac{1}{\beta
^{2}}}\sin \left[ \frac{1}{3}\arcsin \left( \kappa \right) \right] \right]
^{2} \\
\text{with }\kappa &=&\left[ -2+9(-1+3\gamma )\beta ^{2}\right] /\left[
2(1+3\beta ^{2})^{3/2}\right] .
\end{eqnarray*}%
} exists \ for $0<\gamma <1$.

\subsection{ Two-state regime\label{two}}

\subsubsection{Number of solutions\label{NSOL}}

\begin{figure}[t]
\begin{tabular}{cc}
\includegraphics[width=.47\linewidth]{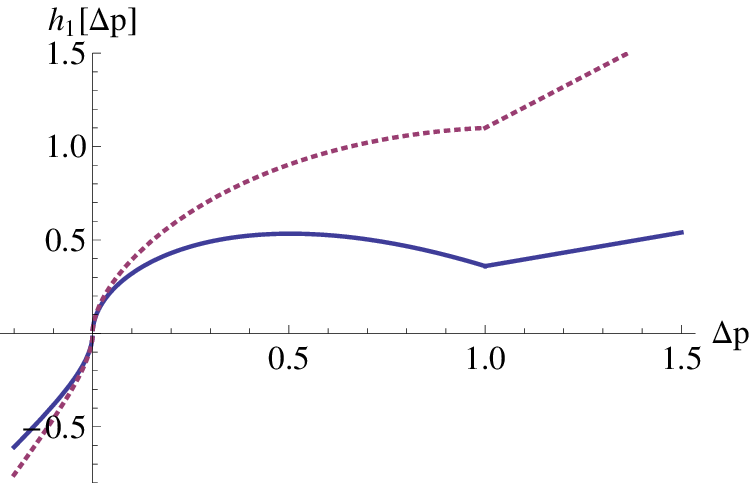} & %
\includegraphics[width=.47\linewidth]{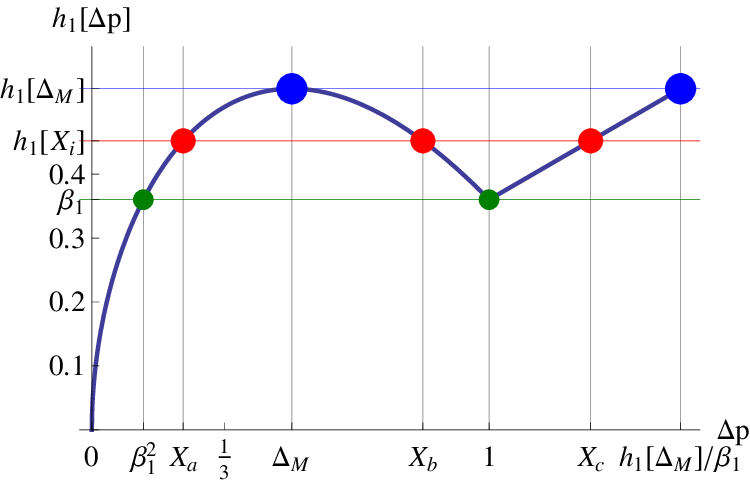}%
\end{tabular}%
\par
\caption{Left: Function $h_{1}(\Delta p)$ for two values of $\protect\beta %
_{1}$. Dotted line: $\protect\beta _{1}=1.1$ (monotonous variation). Solid
line: $\protect\beta _{1}=0.36$. Right: Zoom for the second case. A maximum
value exists at $P_{M}=(\Delta _{M},$ $\ h(\Delta _{M})).$ The middle
horizontal line corresponds to $h_{1}(X)=0.452$, with the 3 solutions $X_{a}$%
, $X_{b}$, $X_{c}.\ $The two other horizontal lines exhibit the limits of
existence of the two-state regime. }
\label{fighx}
\end{figure}

For the two-state regime, two values of $\Delta p$, with the same value of
the function $h_{1}(X)$ are sought. They do not depend on the impedance at
zero frequency, i.e. on the value of $\beta .$ The study of $h_{1}(X)$ leads
to the following results (see Fig. \ref{fighx}): for negative $X$, the
derivative $h_{1}^{\prime }(X)$ is always positive, while for positive $X$,
it is positive up to $X=\Delta _{M}$, where $\sqrt{X}$ is the positive root
of the following equation:%
\begin{eqnarray}
&&3X-2\beta _{1}\sqrt{X}-1=0,\text{ \ i.e.}  \label{e914} \\
&&X=\Delta _{M}\text{ \ with \ }\Delta _{M}=\frac{1}{9}\left[ \beta _{1}+%
\sqrt{\beta _{1}^{2}+3}\right] ^{2}.  \label{e912}
\end{eqnarray}%
For $X>1$ (beating reed), the value of the derivative is $\beta _{1}$, and
is positive. Two cases need to be distinguished: if $\Delta _{M}>1$, i.e. if
$\beta _{1}>1$, the derivative is always positive and it is impossible to
find two values of $X$ with the same value $h_{1}(X).$ On the contrary, for
\begin{equation}
\beta _{1}<1,  \label{e913}
\end{equation}%
the function decreases from $h_{1}(\Delta _{M})$ to $\beta _{1}$ when $X$
increases from $\Delta _{M}$ to $1$, then re-increases. Solutions of Eq. (%
\ref{e889}) can be found in this case only, and the corresponding value of
the function is necessarily larger than $\beta _{1}$. A consequence is that
Eq. (\ref{e889}) has no solution for $h_{1}(X)<\beta _{1}$, and this is true
in particular for negative $h_{1}(X)$.\ Therefore no two-state regime can be
found with negative flow. This conclusion is compatible with the general
result obtained for all possible regimes in Ref. \cite{taillard}. Because of
this result, the paper is now focused on positive pressure differences $%
\Delta p.$

For $h_{1}(X)\in \left[ \beta _{1},h_{1}(\Delta _{M})\right] $, three values
of $X$, defined as $X_{a}<X_{b}<X_{c}$ lead to the same value of the
function (see Fig. \ref{fighx}), and three two-state regimes are possible,
which can be either non-beating (for the pair ($X_{a}$, $X_{b}$)) or beating
(for the pairs ($X_{a}$, $X_{c}$) and ($X_{b}$, $X_{c}$)). The intervals of
the three solutions are as follows: $X_{a}\in \left[ \beta _{1}^{2},\Delta
_{M}\right] $ ; $X_{b}\in \left[ \Delta _{M},1\right] $ ; $X_{c}\geq 1$.\

From the solution of Eq. (\ref{e889}), the corresponding excitation pressure
$\gamma $ for the two-state regime is given by Eq. (\ref{e921a}), for the
three pairs of solutions $(\Delta p_{i},\Delta p_{j})=(X_{i},X_{j})$. For
the particular case of the beating-reed regime, when $h_{1}=\beta _{1}\Delta
p_{c}$, and \ $i=a,b,$ the expression can be also written as:%
\begin{eqnarray}
\gamma _{ic} &=&\frac{1}{2}\left[ h(\Delta p_{i})+\frac{h_{1}(\Delta p_{i})}{%
\beta _{1}}\right]   \label{e92a} \\
&=&\Delta p_{i}+\frac{f(\Delta p_{i})}{\beta _{2}}\overset{def}{=}%
h_{2}(\Delta p_{i})\text{ }.\text{ }
\end{eqnarray}
\subsubsection{ Existence, beating, saturation and extinction thresholds
\label{EB}}

\begin{itemize}
\item At the limit of existence, when $h_{1}^{\prime }(\Delta p)$ vanishes,
\begin{equation}
\Delta p_{a}=\Delta p_{b}=\Delta _{M}.  \label{e923b}
\end{equation}%
Therefore the existence threshold $\gamma _{th}$ of the two-state regime is:
\begin{equation}
\gamma _{th}=\Delta _{M}+\beta f(\Delta _{M})=h(\Delta _{M})  \label{e923}
\end{equation}

The solutions can be either stable or unstable, i.e. the bifurcation can be
direct or inverse.\ This is discussed hereafter in section \ref{sb}.

\item Another limit of existence is given by $X_{b}=1.$ Using, Eq. (\ref%
{e92a}), this yields: $\gamma =\gamma _{c}=1.$ The solutions can be either
stable or unstable, i.e. the bifurcation can be direct or inverse.\ This is
discussed hereafter in section \ref{brca}.

\item The beating threshold $\gamma _{b}$ appears when one solution is $X=1,$
$h_{1}(X)=\beta _{1}$, thus the pair of solution is: $X_{a}=\beta _{1}^{2},$
$X_{b}=1;$ therefore, using Eq. (\ref{e92a}):%
\begin{equation}
\gamma _{b}=h_{2}(\beta _{1}^{2}).  \label{e19}
\end{equation}%
Because $\beta _{1}^{2}<1$, and $0<\beta <\beta _{1}$, the threshold $\gamma
_{b}$ can be shown to be always smaller than the closure threshold $\gamma
_{c}=1$.

\item
The saturation threshold is obtained for the beating regime, with $%
dp_{a}/d\gamma =0,$ with $p_{a}=\gamma -X_{a}$, i.e. $dX_{a}/d\gamma =1$,
therefore $X_{a}=1/3$, $\gamma _{sat}=h_{2}(1/3),$ and%
\begin{equation}
\text{ \ }p_{a}=\frac{2}{3\sqrt{3}\beta _{2}};\text{ }u=f\left( \frac{1}{3}%
\right) \text{ =}\frac{2\zeta }{3\sqrt{3}}=u_{\max }  \label{e18h}
\end{equation}
At the saturation value, the flow rate is maximum: $u=u_{\max }$. For $\beta
_{1}>1/\sqrt{3},$ the saturation threshold is the beating threshold, because
for $\beta _{1}^{2}=1/3$, $\gamma _{sat}=\gamma _{b}$: the amplitude of the
pressure decreases from the beating threshold.

\item Finally the overcritical (extinction) threshold $\gamma _{e}$ is given
by $d\gamma /d\Delta p_{i}=0$ in Eq. (\ref{e92a}). This condition yields to
the following result:%
\begin{eqnarray}
&&3\Delta p_{a}-2\beta _{2}\sqrt{\Delta p_{a}}-1=0,\text{ or }  \label{e18f}
\\
\Delta p_{a} &=&\frac{1}{9}\left[ \beta _{2}+\sqrt{\beta _{2}^{2}+3}\right]
^{2}\text{ }=\Delta p_{e}\text{ }  \label{e18q} \\
&&\text{and }\gamma _{e}\overset{def}{=}h_{2}(\Delta p_{e}).
\end{eqnarray}%
Because $\Delta p_{e}<1$, it exists if $\beta _{2}<1$ only. As explained in
Ref. \cite{Dalmont:05}, when losses tend to $0$ ($\beta _{2}$ tends to $0$),
the extinction threshold tends to infinity. The threshold $\gamma _{e}$ is
always larger than $\gamma _{c}=1$, because $\sqrt{\Delta p_{e}}-\beta
_{2}=(1-\Delta p_{e})/(2\sqrt{\Delta p_{e}})$, thus:%
\begin{equation}
\gamma _{e}-\gamma _{c}=\frac{1}{\beta _{2}}\frac{(1-\Delta p_{e})^{2}}{2%
\sqrt{\Delta p_{e}}}\geq 0.
\end{equation}
\end{itemize}

\subsubsection{Subcritical threshold at emergence (non-beating reed)}

Results (\ref{e923}) to (\ref{e18q}) were obtained with other, equivalent
methods in Ref. \cite{Dalmont:05}. However another threshold can exist for
the non-beating case ($i=a$, $j=b$): for certain values of the parameters,
the emergence bifurcation can be inverse, and the threshold of oscillation
is different for crescendo and decrescendo playing. The subcritical
threshold $\gamma _{sc}$ can be calculated by using the change in variables
defined in Ref. \cite{Dalmont:05}:%
\begin{equation}
\Sigma \overset{def}{=}\sqrt{X_{a}}+\sqrt{X_{b}}\text{ ; }\Pi \overset{def}{=%
}\sqrt{X_{a}}\sqrt{X_{b}}.  \label{AAA}
\end{equation}%
Eq. (\ref{e889}) implies:%
\begin{equation}
\Pi =\Sigma ^{2}-\beta _{1}\Sigma -1  \label{AZR}
\end{equation}%
(this change in variables is related to the decomposition into dc and
acoustic components, see Eq. (\ref{e887})). The threshold can be calculated
by writing $d\gamma /d\Sigma =0$ in Eq. (\ref{e921a}), which can be written
as a polynomial in $\Sigma $ (see Eq. (A17) in Ref. \cite{Dalmont:05})%
\footnote{%
The equation is:
\begin{equation*}
\gamma =\beta \Sigma ^{3}-(1+3\beta \beta _{1})\Sigma ^{2}/2+(\beta
_{1}-\beta )\Sigma +1.
\end{equation*}%
The two solutions for $\varphi $ is as follows (see Eq. (\ref{e65a}):%
\begin{eqnarray*}
\varphi _{n} &=&\frac{1}{3}\left( \beta _{3}+2\delta \cos \left[ \frac{1}{3}%
\left( \arccos (\Phi )+2n\pi \right) \right] \right)  \\
\text{where }\Phi  &=&\frac{2\beta \beta _{3}\left( \delta ^{2}-12\right)
+27\left( \gamma -\gamma _{b}\right) }{2\beta \delta ^{3}} \\
\text{and }\delta  &=&\sqrt{3+\beta _{3}^{2}}\text{ ; }n=0\text{ or }2.
\end{eqnarray*}%
\par
}$.$ For our
purpose, it is convenient to write this equation as follows (denoting $%
\gamma =\gamma _{ab})$:%
\begin{equation}
\gamma =\beta (\varphi -\beta _{3})(\varphi ^{2}-1)+\gamma _{b}\text{ \ with
}\varphi \overset{def}{=}\Sigma -\beta _{1},  \label{A}
\end{equation}%
where $\beta _{3}$ is given by Eq. (\ref{e65a}). Notice that $\varphi
_{b}=1. $ The derivative $d\gamma /d\varphi $ vanishes for $\varphi =\varphi
_{sc}$:%
\begin{eqnarray}
3\varphi _{sc}^{2}-2\beta _{3}\varphi _{sc}-1 &=&0\text{ \ or }  \label{CZ}
\\
2\varphi _{sc}(\varphi _{sc}-\beta _{3}) &=&1-\varphi _{sc}^{2}\text{ \ \ \
or }  \label{BZ} \\
\varphi _{sc} &=&\frac{1}{3}\left[ \beta _{3}+\sqrt{\beta _{3}^{2}+3}\right]
.  \label{CY}
\end{eqnarray}%
Therefore, $\varphi _{sc}$ is always positive, and combining Eqs. (\ref{A})
and (\ref{BZ}), it is shown that the threshold $\gamma _{sc}$ is always
smaller than the beating-reed threshold:%
\begin{equation}
\gamma _{sc}=\gamma _{b}-\frac{\beta }{2\varphi _{sc}}(1-\varphi
_{sc}^{2})^{2}.  \label{E}
\end{equation}

\subsubsection{Solutions}

The direct solution of the cubic equation (\ref{A}) is possible (see
footnote 4). In the present paper we propose a method based upon Eqs. (\ref%
{e889}) and (\ref{e921}). Starting from a value of $X_{a}\in \left[ \beta
_{1}^{2},\Delta _{M}\right] $, the solution $X_{c}$ (above unity) is
obtained by $X_{c}=h(X_{a})/\beta _{1}$, and the solution $X_{b}$ is deduced
by solving the equation $h_{1}(X_{b})=h_{1}(X_{a})$ for $X_{b}\in $ $\left[
\Delta _{M},1\right] $. The latter equation is cubic in $\sqrt{X_{b}}$. It
has a solution already known, $\sqrt{X_{a}}$, therefore $\sqrt{X_{b}}$ can
be deduced as the positive solution of a quadratic equation\footnote{%
According to Vieta's formula, the sum of the three solutions is $\beta _{1}$%
, their product is $-h.$ Notice that the third solution differs from $\sqrt{%
X_{c}}$, because $X_{c}$ corresponds to a beating reed.}, as follows:%
\begin{equation*}
x^{2}-(\beta _{1}-\sqrt{X_{a}})x-h_{1}(X_{a})/\sqrt{X_{a}}=0\text{, where }x=%
\sqrt{X_{b}}.
\end{equation*}%
\begin{equation}
X_{b}=\frac{1}{4}\left[ \beta _{1}-\sqrt{X_{a}}+\sqrt{(\beta _{1}+\sqrt{X_{a}%
})^{2}+\frac{4f(X_{a})}{\sqrt{X_{a}}}}\,\right] ^{2}.  \label{e926}
\end{equation}%
From the knowledge of the three solutions of $h_{1}(X)=h_{1}(X_{a})$, the
three values of $\gamma _{ij}$ are deduced from Eq. (\ref{e921}). Figures \ref%
{pfigA} and \ref{pfigB} show 4 examples of bifurcation schemes for different cases.  The calculation is done by varying the starting value $X_{a}$  and was verified using the
iterated map algorithm (Ref. \cite{taillard}), which obviously gives stable
solutions only.
The three straight lines $p=\gamma -\beta _{1}^{2}$, $p=\gamma
-\Delta _{M}$, and $p=\gamma -1$ delimit the three domains of solutions, $%
X_{a}$, $X_{b}$ and $X_{c}.$ The solution is non-beating for the pair ($%
X_{a}$, $X_{b}$)) or beating, for the pairs ($X_{a}$, $X_{c}$) and ($X_{b}$,
$X_{c}$)), see Fig. 1.
\onecolumn
\begin{figure}
\begin{tabular}{cc}
\includegraphics[width=.5\linewidth]{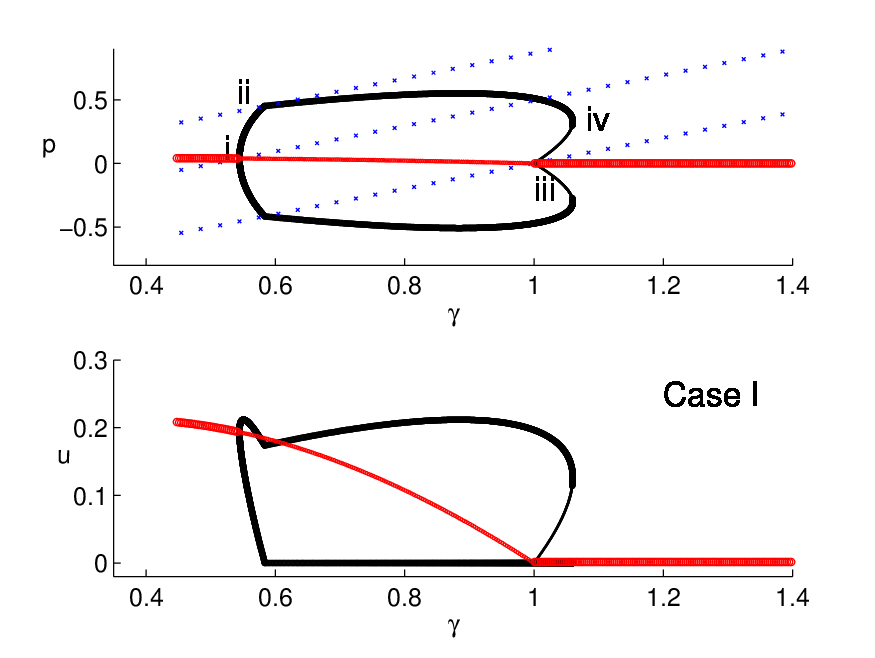} & %
\includegraphics[width=.5\linewidth]{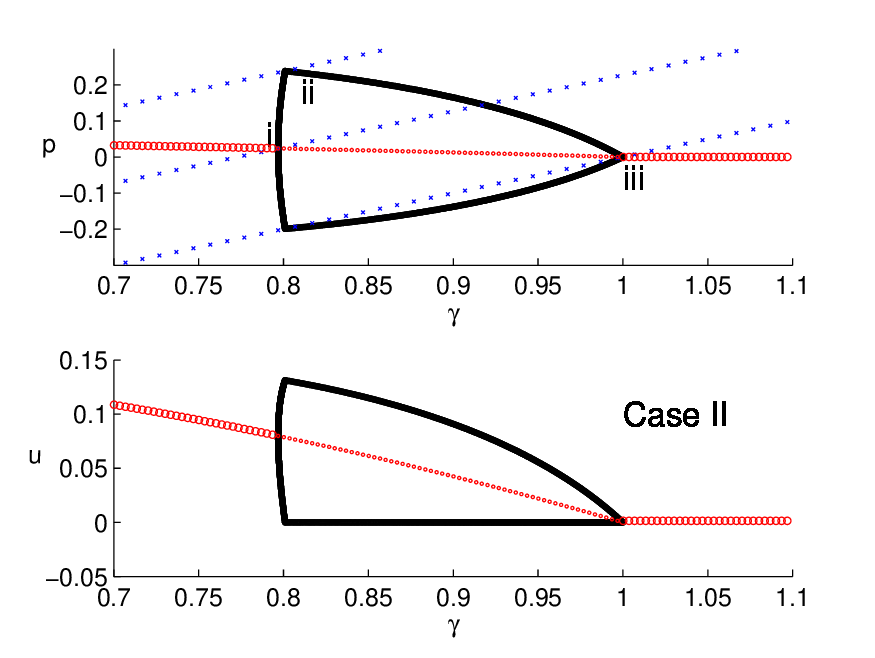}
\end{tabular}
\caption{Two examples of bifurcation schemes for the pressure $\gamma ,p$ and the flow rate $\gamma ,u$. The emergence bifurcation is direct. The three dotted (blue), increasing straight lines indicate the three domains of solutions $%
X_{a},$ $X_{b}$, and $X_{c}$. \textbf{Case I, left}: $\protect\zeta =0.55 $, $\protect\mu =0.2$.%
The solid curve corresponds to a two-state regime, which is stable (thick line) or unstable (thin line):
the emergence bifurcation is direct, while the extinction bifurcation is
inverse. The pale (red), decreasing line corresponds to the static regime,
which is stable (thick line) or unstable (thin line).  The point $i$ indicates the instability threshold $%
\protect\gamma _{th}$ of the static regime; the point $ii$ the beating
threshold $\protect\gamma _{b}$ of the two-state regime; the point $iii$ is
the closure threshold $\protect\gamma _{c}=1$, \ the point $iv$ is the
extinction threshold $\protect\gamma _{e}$ of the two-state regime. \textbf{Case II, right}: $\protect\zeta =0.4$, $\protect\mu =0.3$.  Both emergence $i$ and extinction bifurcations $iii$ are direct. The point $ii$ indicates the beating threshold $\protect\gamma _{b}$ of the two-state regime. Because $\protect\beta _{1}=.75>1/\protect\sqrt{3}$, saturation occurs at the beating threshold $\protect\gamma _{b}.$}
\label{pfigA}
\end{figure}

\begin{figure}
\begin{tabular}{cc}
\includegraphics[width=.5\linewidth]{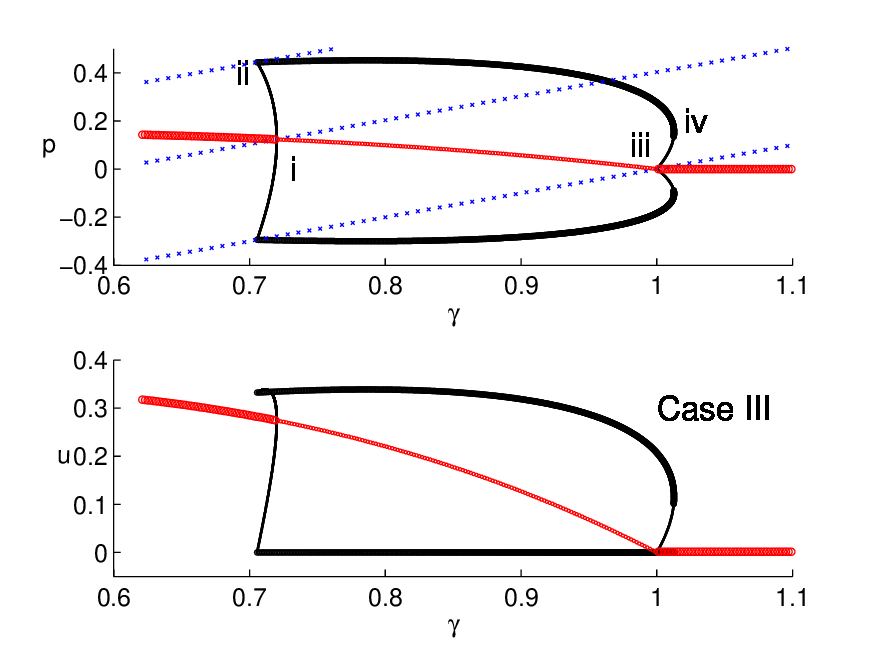} & %
\includegraphics[width=.5\linewidth]{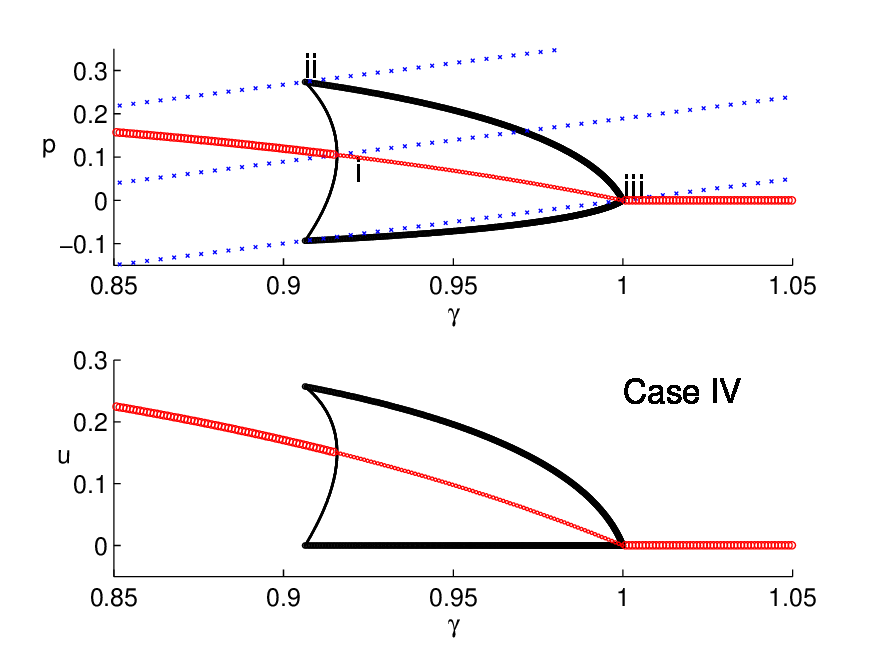}
\end{tabular}
\caption{Two examples of bifurcation schemes for the pressure $\gamma ,p$ and the flow rate $\gamma ,u$. The emergence bifurcation is inverse. The three dotted
(blue), increasing straight lines indicate the three domains of solutions $%
X_{a},$ $X_{b}$, and $X_{c}$. \textbf{Case III, left}: $\protect\zeta =0.88$, $\protect%
\mu =0.45$. Both emergence and extinction bifurcations
are inverse. The point $iv$ indicates the overcritical (extinction)
threshold $\protect\gamma _{e}$ of the two-state regime, and the point $iii$
the closure threshold $\protect\gamma _{c}=1$ of the static regime. \textbf{Case IV, right}: $\protect\zeta =0.88$, $\protect\mu =0.7$.  The extinction bifurcation
 is direct.  }
\label{pfigB}
\end{figure}

\twocolumn

\begin{figure}[t]

 \includegraphics[width=.9\linewidth]{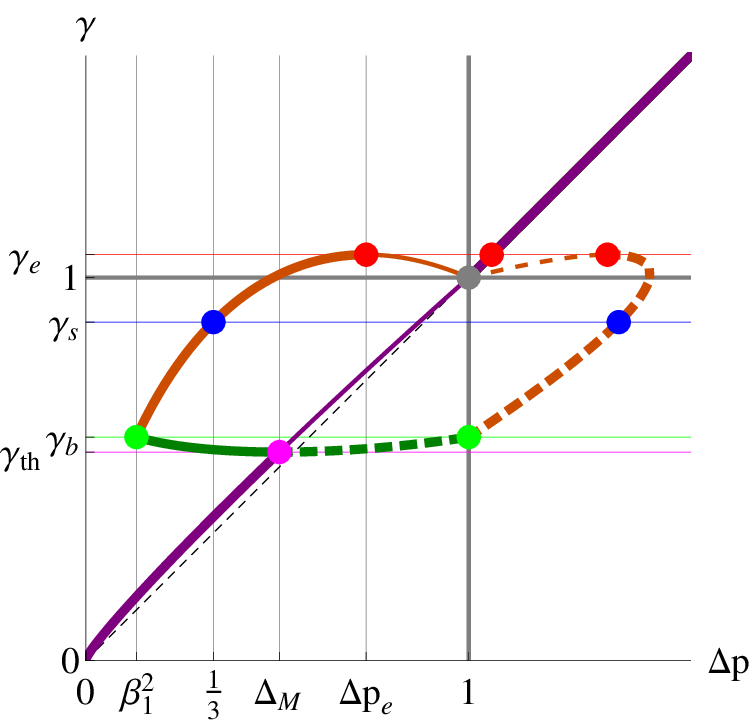}%

\caption{
Representation for $\Delta p$\textbf{,}$\protect\gamma )$\textbf{.}
The function $h(\Delta p)$ is used for the static regime (black curve,
purple in color) and the function $h_{2}(\Delta p)$ for the beating,
two-state regime (grey curve, brown in color). The non-beating two-state
(light grey, green in color) is given by Eqs. (\protect\ref{e921a}) and (%
\protect\ref{e926}). Thick lines: stable regime, thin lines: unstable
regime. Dashed lines: solution of the 2-state regime with the high value of $%
\Delta p$. The points signal the transitions at the different thresholds.}
\label{pfigIbis}
\end{figure}

Figure \ref{pfigIbis} shows the diagram (%
$\Delta p,\gamma $) for the case I of Figure \ref{pfigA}.   This allows exhibiting the function $h(\Delta p),$
given by Eq. (\ref{e8}), for the static regime and the function $%
h_{2}(\Delta p)$, given by Eq. (\ref{e92a}), for the beating, two-state
regime. The one-state regime is stable from $\{0,0\}$ to $\{\Delta
_{M},\gamma _{th}\}$ and above $\{1,1\}$. The non-beating two-state is given
by Eqs. (\ref{e921a}) and (\ref{e926}), the beating two-state is given for $%
X_{a}<1$ by the function $h_{2}(X)$. The bifurcation at emergence is direct,
the oscillation is stable from $\{\Delta _{M},\gamma _{th}\}$ to $\{\beta
_{1}^{2},\gamma _{b}\}$. The beating two-state is stable from $\{\beta
_{1}^{2},\gamma _{b}\}$ to $\{\Delta p_{e},\gamma _{e}\}$ and unstable
between $\{\Delta p_{e},\gamma _{e}\}$ and $\{1,1\}$. The oscillation
amplitude is at a maximum at $\{1/3,\gamma _{s}\}$, where the slope of $%
h_{2}(X)$ is unity.

\subsubsection{Radiated sound\label{RS}}

The radiation losses are small at low frequencies, therefore it is very
simple and classical to deduce them by perturbation from the output flow
rate, considering a monopole radiation. Two cases have to be distinguished:
losses occur at the output, or losses occur during propagation into the
tube. Both cases give the same input impedance, but not the same transfer
functions between input and output. The latter case is more realistic, and
is considered here. If the output impedance is $0,$ and losses due to
boundary layers only, the output acoustic flow rate is given by the standard
transmission lines relationships: $u_{out}=\sinh (j\omega \ell /c+\alpha
\ell )p_{ac}$, where $\omega \ell /c=\pi /2$ (see section \ref{besr}).
Therefore the amplitude relationship is the following:%
\begin{eqnarray*}
\left\vert u_{out}\right\vert  &=&\left\vert p_{ac}\right\vert /\cosh
(\alpha \ell )= \\
\left\vert p_{ac}\right\vert \sqrt{1-\mu ^{2}} &=&\left\vert
u_{ac}\right\vert \frac{\sqrt{1-\mu ^{2}}}{\mu }.
\end{eqnarray*}%
The maximum output acoustic flow rate is obtained for the saturation
threshold; at the saturation threshold, because the reed is beating, $%
\left\vert u_{ac}\right\vert =(u_{\max }+0)/2=\zeta /(3\sqrt{3}),$(see Eq. (%
\ref{e18h})). For $\beta _{1}=\mu /\zeta <1/\sqrt{3},$ the saturation
threshold is the beating threshold (see section \ref{EB}), and $%
p_{ac}=(1-\beta _{1}^{2})/2$, therefore:%
\begin{eqnarray}
\left\vert u_{out}\right\vert _{\max } &=&\frac{\sqrt{1-\mu ^{2}}}{3\sqrt{3}%
\beta _{1}}\text{ if }\beta _{1}>\frac{1}{\sqrt{3}};  \label{rad} \\
\left\vert u_{out}\right\vert _{\max } &=&\frac{1}{2}(1-\beta _{1}^{2})\sqrt{%
1-\mu ^{2}}\text{if }\beta _{1}<\frac{1}{\sqrt{3}}.  \notag
\end{eqnarray}%
For a given value of $\zeta $, this is monotonously decreasing function of $%
\mu $; for a given value of $\mu $, this is a increasing function of $\zeta .
$

\section{Calculation of instability thresholds \ \label{CT}}

The stability conditions (\ref{e10}) and (\ref{e301}) are calculated using
the expression of the derivative $f^{\prime }(\Delta p)$:%
\begin{eqnarray}
\text{ }f^{\prime }(\Delta p) &=&\frac{1-3\Delta p}{2\sqrt{\Delta p}}\text{
\ if }\Delta p<1\text{ ;}  \notag \\
\text{ }f^{\prime }(\Delta p) &=&0\text{ if }\Delta p>1.
\end{eqnarray}

\subsection{Instability of the static regime\label{is}}

The condition (\ref{e10}) generalizes the condition $f^{\prime }(\Delta p)>0$%
, as discussed in Ref. \cite{flet} for the lossless case (see page 349). For
the (static) beating reed case, which exists for $\gamma >\gamma _{c}=1$
(see previous section), $f^{\prime }(\Delta p)=0$, thus the static regime is
always stable. For the non-beating reed case, the first factor of Inequality
(\ref{e10}) is equal to $h_{1}^{\prime }(\Delta p)$, thus it vanishes when $%
\Delta p$ satisfies Eq. (\ref{e914}), i.e. $\Delta p=\Delta _{M}$. For this
value, $\Delta _{M}$, the second factor of Inequality (\ref{e10}) is $%
(1-\beta \beta _{1})$, which is positive\footnote{%
Another threshold can be sought when the denominator of Eq. (\ref{e10})
vanishes: this leads to the solution of either $1=0$ when $\Delta p>1$, or $%
f^{\prime }(\Delta p)=-1/\beta $, with $0<\Delta p<1$. These two equations
have no solution.}. Thus (\ref{e912}) together with Eqs. (\ref{e8}) gives
the threshold $\gamma _{th}$, given by Eq. (\ref{e923}).

Therefore the condition $h_{1}^{\prime }(\Delta p)=0$ gives both the limit
of existence of the two-state regime (see Eq. (\ref{e923})), and the
instability threshold of the static regime.\ Nevertheless the nature (direct
or inverse) of the bifurcation between the two regimes is not yet solved by
this result\footnote{%
In Ref. \cite{Dalmont:05}, $\beta $ was assumed to be very small in
practice, and Eq. (\ref{e923}) was simplified in $\gamma _{th}=\Delta _{M}$,
but the complete Eq. (\ref{e923}) was given for the threshold of existence
for the two-state regime.}. The term $\beta f(\Delta _{M})$ is the static
pressure $p_{s}$ \ in the mouthpiece; the presence of the parameter $\beta $
indicates that the threshold depends on the impedance at zero frequency.

\subsection{Instability of the two-state regime\label{ET2}}

\subsubsection{\label{brca} Beating-reed case: overcritical (extinction)
threshold}

For the beating regime $f^{\prime }(X_{c})=0$, with $X_{i}<1,$ thus, using
Condition (\ref{e301}), $C=-f^{\prime }(X_{i})$ \ for $i=a$ or $b$.\ Because
$C<1$, and because $\zeta ^{2}\beta _{2}=\zeta \tanh 2\alpha \ell <1$, the
second inequality (\ref{e301}) is never valid. Therefore the stability is
defined by the condition $C<\beta _{2}.$

When $\beta _{2}>1$, this condition is always satisfied, and the beating
two-state regime is stable (no overcritical threshold $\gamma _{e}$ exists,
as noticed in Section \ref{EB}).

When $\beta _{2}<1$, Eq. (\ref{e18f}) is used, yielding $\beta _{2}=(3\Delta
p_{e}-1)/(2\sqrt{\Delta p_{e}})$. The function $(3x-1)/(2\sqrt{x})$ being
always increasing for positive $x$, the inequality $C<\beta _{2}$ holds if $%
X_{i}<\Delta p_{e}$, or $p_{i}>p_{e}=\gamma -\Delta p_{e}.$ This
distinguishes in the ($\gamma ,p$) plane the two branches separated by the
overcritical threshold: the upper one is stable, while the lower one is
unstable\footnote{%
It is possible to show that the interesting solution in this discussion is $%
X_{i}=X_{b}$: because $\beta _{2}>\beta _{1}$, the solution $\Delta p_{e}$
at the overcritical threshold is larger than $\Delta _{M}$, thus it is
always larger than $X_{a}.$ Instability occurs for the pair ($X_{b},$ $X_{c}$%
).}. The unstable
branch is the branch joining the static regime, because the two regimes
cannot be stable for the same value of the parameter $\gamma $ when they
converge to the same point. This can be explained with mathematical
arguments, based upon either a perturbation method (see Ref. \cite{115}) or
the topological degree (see Ref. \cite{171}). This can be also studied by
using Inequalities (\ref{e301}), as done in Ref. \cite{Dalmont:05}.

It can be noticed that because $\gamma _{e}$ is larger than unity, the
beating, two-state regime is always stable for $\gamma _{b}<\gamma <\gamma
_{c}=1$, whatever the value of $\beta _{2}.$

\subsubsection{Non-beating case: period doubling and subcritical (emergence)
threshold \label{sb}}

i) For the non-beating regime ($i=a$, $j=b$), an expression of the
instability threshold was given in Ref. \cite{Dalmont:05} and it was
explained that the threshold is given by $1/C=\zeta ^{2}\beta _{2}$. Some
errors were done in the derivation, and they are corrected in appendix A of
the present paper. When the excitation pressure is larger than this
threshold, here denoted $\gamma _{ins}$, period doubling can occur, then a
complex bifurcation scenario (see Ref. \cite{taillard}).

ii) Otherwise, it can be checked that the second condition $C=\beta _{2}$
leads to the subcritical emergence threshold $\gamma _{sc}$: it separates
two branches in the ($\gamma ,p)$ plane, the upper one being stable while
the lower one is unstable. This is similar to what happens for the
overcritical threshold. When it exists, the emergence bifurcation is as
follows: when playing crescendo, the oscillation starts for $\gamma =\gamma
_{th}$, while playing decrescendo, the oscillation stops for $\gamma =\gamma
_{sc}.$

When this subcritical threshold can exist? From Eq. (\ref{A}) and (\ref{CZ}%
), it is found that
\begin{equation}
\gamma _{th}-\gamma _{sc}=\beta (\varphi _{th}-\varphi _{sc})^{2}\left[
\frac{1}{2\varphi _{sc}}+\frac{\varphi _{sc}}{2}+\varphi _{th}\right]
\label{CQ}
\end{equation}%
where $\varphi _{th}=\Sigma _{th}-\beta _{1}$ and $\varphi _{th}-\varphi
_{sc}=\Sigma _{th}-\Sigma _{sc}$, with
\begin{equation}
\text{ }\Sigma _{th}=2\sqrt{\Delta _{M}}=\frac{2}{3}\left( \beta _{1}+\sqrt{%
\beta _{1}^{2}+3}\right) \text{.}  \label{504}
\end{equation}

Two cases are possible:

\begin{itemize}
\item $\Sigma _{sc}>\Sigma _{th}$: the emergence bifurcation is direct:
stable solutions exist for $\gamma >\gamma _{th}$.

\item $\Sigma _{sc}<\Sigma _{th}$: the emergence bifurcation is inverse, and
stable solutions exist for $\gamma >\gamma _{sc}.$
\end{itemize}

When $\Sigma _{sc}$ continues to decrease below $\Sigma _{th}$, the
bifurcation remains inverse, but the subcritical threshold $\gamma _{sc}$
becomes the beating threshold $\gamma _{b}$. This happens when $\Sigma
_{sc}=\Sigma _{b}=1+\beta _{1}$, or $\varphi _{sc}=\varphi _{b}=1$ (notice
that the inequality $\Sigma _{b}\leq \Sigma _{th}$ always holds). The
discussion is extended in section \ref{OL}.

\section{Limits of regimes in the plane ($\protect\mu ,\protect\zeta )$ \
\label{DIA}}

\begin{figure}[t]
\begin{center}
\includegraphics[width=8cm]{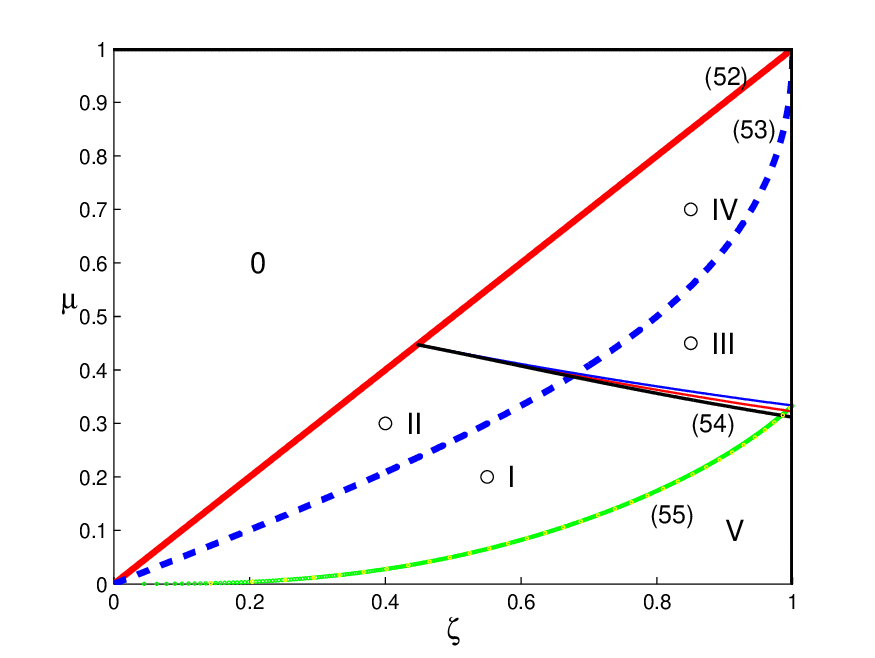}
\end{center}
\caption{{}Plane $(\protect\zeta ,\protect\mu ).$ The four Regions I to IV
correspond to the different cases presented for the examples shown in Figs.
 \ref%
{pfigA} and \ref{pfigB}. The
chosen values for the examples are indicated by a circle. In Region V period
doubling or other regimes can exist, in Region 0 no sound is possible. The
numbers refer to the equation number. Above line (\protect\ref{e22}), no
sound is possible. Curve (\protect\ref{e23}) distinguishes the extinction
bifurcation (direct above, inverse below). Curve (\protect\ref{e24})
distinguishes the emergence bifurcation (direct above, inverse below). Below
curve (\protect\ref{e237}), complicated regimes can appear by
destabilization of the two-state regime. The two curves close to the curve (%
\protect\ref{e24}) correspond to equations (\protect\ref{e25}), and (\protect
\ref{e236}), from to the top to the bottom (see section \protect\ref{OL}).}
\label{pertes4}
\end{figure}

From the different expressions of the thresholds, it is possible to deduce
the limits separating different domains in the plane $(\mu ,\zeta )$, as
shown in Fig. \ref{pertes4}. Above the diagonal (region 0), no two-state
regime can exist. Other regions of the plane are defined by the nature of
the emergence and extinction bifurcations: they are named by the number of the four cases shown in
Figures Figures \ref%
{pfigA} and \ref{pfigB}: in Regions I and III, the
extinction bifurcation is inverse, while in Regions II\ and IV, it is
direct.\ What is new in this paper is the separation between Regions I and
II, with direct emergence bifurcation, and regions III and IV, with inverse
emergence bifurcation. Finally, in Region V, the two-state regime can be
unstable, and can be replaced by more complicated regimes, with period
doubling, chaos, intermittences, etc... (see Ref. \cite{taillard}).

\subsection{Emergence and extinction bifurcations}

\begin{itemize}
\item When the instability threshold $\gamma _{th\text{ }}$(Eq. (\ref{e923}%
)) of the static regime reaches the closing threshold $\gamma _{c}=1$, the
static regime becomes stable whatever the values of all parameters, and no
sound can be expected. This happens\footnote{%
The Eq. $\gamma _{th}=1$ has two solutions: $\Delta _{M}=1$, and $\sqrt{%
\Delta _{M}}=1/\beta $. Because $\beta $ is necessarily less than unity, the
latter solution is larger than unity.} for $\beta _{1}>1$, and this confirms
the result explained in section \ref{NSOL} that no two-state regime can
exist (this discussion is extended in the next section). The condition for
the existence of sound can be also written as:
\begin{equation}
\mu <\zeta \text{, or \ \ }\lambda >\frac{1-\zeta }{1+\zeta }\text{ .}
\label{e22}
\end{equation}

\item When the overcritical threshold $\gamma _{e}$ (Eq. (\ref{e18q}))
reaches the closure threshold of the static regime $\gamma _{c}=1$ (Eq. (\ref%
{e19})), the extinction bifurcation becomes direct instead of inverse, as
explained in the previous section. For $\beta _{2}<1,$ the bifurcation is
inverse: this is probably the most frequent case for real clarinets and
clarinettists, and corresponds to:%
\begin{equation}
\zeta >\tanh 2\alpha \ell =\frac{2\mu }{1+\mu ^{2}},\text{ or \ \ }\lambda >%
\sqrt{\frac{1-\zeta }{1+\zeta }}.  \label{e23}
\end{equation}%
Between the two limits (\ref{e22}) and (\ref{e23}), the extinction
bifurcation is direct (Regions I and IV).

\item When the subcritical threshold $\gamma _{sc}$ (Eq. (\ref{CY})) reaches
the instability threshold of the static regime $\gamma _{th}$ (Eq. (\ref%
{e923}), the emergence bifurcation becomes direct. From Eq. (\ref{CQ}), this
happens when $\varphi _{sc}=\varphi _{th}$. The bifurcation is direct in
Regions I and II, with the following condition:
\begin{equation}
\beta <\frac{\Sigma _{th}-\beta _{1}}{\beta _{1}\Sigma _{th}+3}.  \label{e24}
\end{equation}
\end{itemize}

\subsection{Limit of instability of the two-state, non-beating regime\label%
{LIT}}

Finally, when the instability threshold $\gamma _{ins}$ of the oscillating
regime given by the condition $1/C=\zeta ^{2}\beta _{2}$ (see section \ref%
{sb}) reaches the beating threshold $\gamma _{b}$, the limit was given in
Ref. \cite{Dalmont:05} (with a small error). The formula can be obtained
from Eq. (\ref{e301}), for $X_{a}=\beta _{1}^{2}$, and $X_{b}=1$, $f^{\prime
}(X_{b})=-1.$ The limit $\zeta _{i}$ is given by the following equation:%
\begin{equation}
\frac{1-f^{\prime }}{1-\zeta ^{2}f^{\prime }}=\frac{1}{\zeta ^{2}\beta _{2}}%
\text{ \ where }f^{\prime }=f^{\prime }(X_{a})=\frac{1-3\beta _{1}^{2}}{%
2\beta _{1}}.  \label{e237}
\end{equation}%
This leads to a second order equation in $\zeta ^{2}$:%
\begin{equation}
\beta _{1}^{2}(1-3\beta _{1}^{2})\zeta ^{4}+(4\beta _{1}^{2}-3\beta
_{1}+1)(\beta _{1}+1)\zeta ^{2}-2\beta _{1}=0.  \label{e2365}
\end{equation}%
A particular value is $\beta _{1}=1/3$, $f^{\prime }=1$, $\zeta _{i}=1$, $%
\mu =1/3$, $\lambda =1/2$. For a given loss coefficient $\mu $ $(<\zeta )$,
the two-state regime is always stable when $\zeta <\zeta _{i}$. For sake of
simplicity, we remark that an excellent approximation of the limit, better
than 1\%, is based upon the fact that for small losses, the coefficient $%
\beta =\zeta ^{2}\beta _{1}$ is small, therefore $\beta _{2}\simeq 2\beta
_{1}$:%
\begin{equation}
\zeta _{i}^{2}\simeq \frac{1}{2\beta _{1}(1-f^{\prime })+f^{\prime }}.
\label{e238}
\end{equation}

\subsection{Limits related to the beating threshold\label{OL}}

Another limit is reached when the subcritical threshold $\gamma _{sc}$
becomes the beating-reed threshold $\gamma _{b}$, thus the stable two-state
regime is always beating. However the unstable branch is non-beating (see
Figs. \ref{pfigA} and \ref{pfigB}). Using Eq. (\ref{E}), it is found to
be given by $\varphi _{sc}=\beta _{3}=1$, i.e.%
\begin{equation}
\beta <\frac{1}{3\beta _{1}+2}\text{ \ }.  \label{e25}
\end{equation}

\ A last limit is reached when the instability threshold of the static
regime $\gamma _{th}$ (\ref{e923}) reaches the beating threshold of the
two-state regime $\gamma _{b}$, i.e. when $\varphi _{th}=\Sigma _{th}-\beta
_{1}=\beta _{3}$ in Eq. (\ref{A}):%
\begin{equation}
\beta =\frac{1}{\beta _{1}+2\Sigma _{th}}.  \label{e236}
\end{equation}%
For both limits given by Eqs. (\ref{e25}) and (\ref{e236}), the
corresponding curves are therefore within Regions III and IV of Figure \ref%
{pertes4}. They are very close to the limit given by Eq. (\ref{e24}),
corresponding to the change in nature of the emergence bifurcation. The
three limits are even equal for $\beta _{1}=1$, $\beta =1/5$, $\mu =\zeta =1/%
\sqrt{5}.$ For a given $\zeta $, when losses are small ($\mu $ small), the
emergence bifurcation is direct. Then, when $\beta $ reaches the limit (\ref%
{e24}), the bifurcation becomes inverse, with a non-beating reed. Then, when
$\beta $ reaches the limit given by Eq. (\ref{e25}), the bifurcation remains
inverse, but the reed becomes always beating in the two-state regime.

\begin{figure}[tbp]
\begin{center}
\includegraphics[width=5cm,keepaspectratio=true]{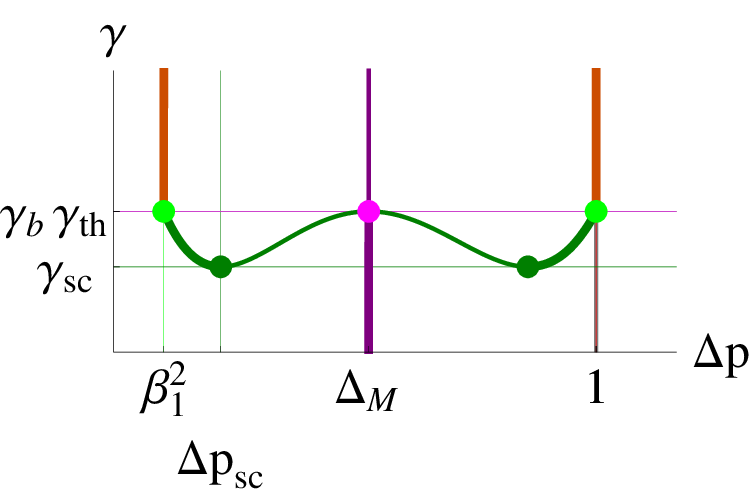} %
\includegraphics[width=5cm,keepaspectratio=true]{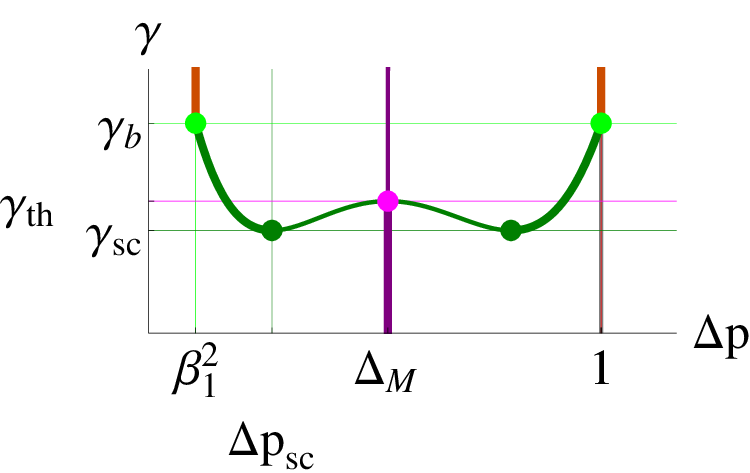} %
\includegraphics[width=5cm,keepaspectratio=true]{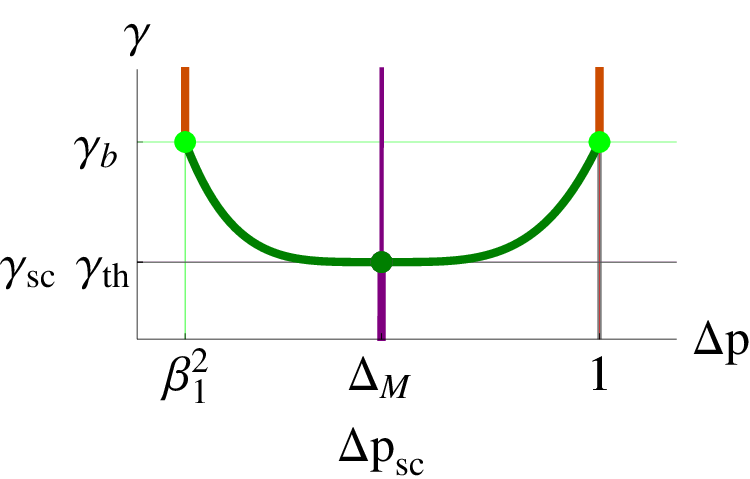} %
\includegraphics[width=5cm,keepaspectratio=true]{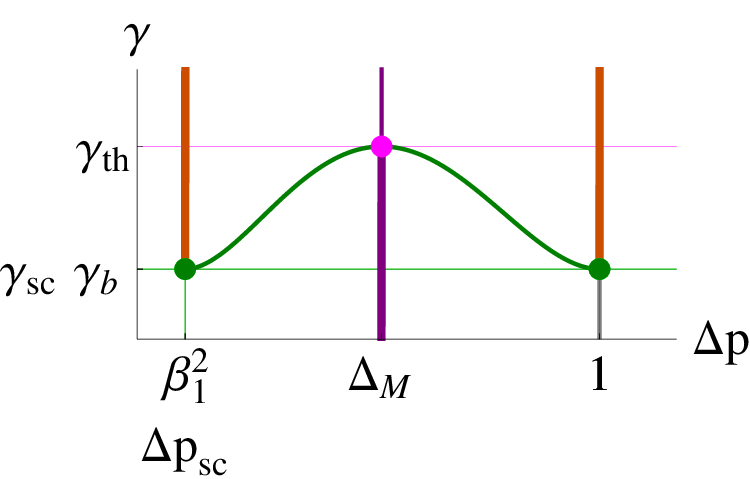}
\end{center}
\par
%\caption{Zoom of bifurcation diagrams with sub-critical threshold. \textbf{%
%Left up:} case $\protect\gamma _{sc}<\protect\gamma _{th}<\protect\gamma %
%_{b}.$ \textbf{Right up:} case $\protect\gamma _{sc}=\protect\gamma _{th}$.
%\textbf{Left down:} case $\protect\gamma _{th}=\protect\gamma _{b}$. \textbf{%
%Left down:} case $\protect\gamma _{sc}=\protect\gamma _{b}$. }
\caption{Zoom of bifurcation diagrams with sub-critical threshold. $\protect%
\beta _{1}=0.68$ \textbf{Left up:} case $\protect\gamma _{sc}=\protect\gamma %
_{th}$, $\protect\beta =0.24584;$ \textbf{Right up:} case $\protect\gamma %
_{sc}<\protect\gamma _{th}<\protect\gamma _{b}$, $\protect\beta =0.24532$;
\textbf{Left down:} case $\protect\gamma _{th}=\protect\gamma _{b}$, $%
\protect\beta =0.24752$ \textbf{Right down:} case $\protect\gamma _{sc}=%
\protect\gamma _{b}$, $\protect\beta =0.24322.$ }
\label{figBdZoom}
\end{figure}

Fig. \ref{figBdZoom} shows details of the bifurcation scheme ($\Delta
p,\gamma )$ near the subcritical threshold for different cases between the
non-beating case $\gamma _{th}<\gamma _{b}$ and the beating case $\gamma
_{b}<\gamma _{th}$ (for the latter case, the beating threshold is the
subcritical threshold, as for the points III and IV). The corresponding
values of the parameter ($\zeta ,\mu $) are extremely close together.

\section{Influence of losses on the existence of the two-state regime\label%
{disc}}

When losses tend to infinity ($\mu $ tends to unity), no sound is possible
whatever the value of $\zeta $: this is in accordance with the intuition
that if no reflection exists at the input of the resonator, no
self-sustained oscillation can happen. Nevertheless, we do not prove that
other types of regimes cannot exist, such as four-state, eight-state, ... In
this section, this issue\ is discussed together with the influence of the
choice of the model. Moreover, some musical consequences are discussed.

\subsection{Possibility of existence of other oscillating regimes}

When the static regime is always stable ($\beta _{1}>1$), it has been proved
that the two-state regime cannot exist. It is probable that other types of
regimes do not exist, but the general proof is difficult. Using the
calculation of the successive iterates functions for different values of the
initial condition (see Ref. \cite{taillard}), we verified that when $\beta
_{1}>1$, the successive iterates converge to the unique point that is the
limit cycle of the static regime, thus no other regime can exist. This can
be done for every set of parameter values, $\gamma $, $\zeta $, $\mu <\zeta $%
: if the convergence is always to a unique state, then it is sure that no
other regime than the static one can exist. Obviously, this verification is
not possible in practice, but the verification has been done for some set of
parameter values.

We conclude that no oscillating regime exists for $\beta _{1}>1$, even if
the rigorous sentence should be: no oscillating regime exists by
destabilization of the static regime. A similar discussion could be done for
the destabilization of the two-state regime into more complicated regimes
(see section \ref{LIT}).

\subsection{ Influence of the choice of the model}

\subsubsection{Model for the beating reed}

The method used in the present paper can also be used for any shape of the
nonlinear characteristic, at least numerically (the condition being that a
nonlinear characteristic is static).\ All equations of sections \ref{model}
and \ref{recu} remain valid when modifying the nonlinear function $f(\Delta
p).$ In particular if a smooth beating transition is chosen with no
singularity, Fig. \ref{fighx} shows that the condition (\ref{e913}), $\beta
_{1}<1$, can be generalized into the following condition: the function $%
h_{1}(\Delta p)$ goes through both a maximum and a minimum.

\subsubsection{Frequency dependence of losses\label{FDL}}

The Raman model is interesting because all quantities can be determined
analytically, but it is not very realistic. It is based upon two important
assumptions: losses do not depend on frequency, and the reed has no
dynamics. A generalization of the present study is out of the scope of the
present paper, but it is interesting to note that the condition $\beta
_{1}<1 $ can be easily generalized when these assumptions are not done, as
explained hereafter.

When losses depend on frequency, it is possible to use the characteristic
equation obtained by linearizing the nonlinear equation around the pressure
of the static regime $p_{s}$, and writing the approximation of the first
harmonic:%
\begin{eqnarray}
u &=&F(p)\simeq F(p_{s})+(p-p_{s})F^{\prime }(p_{s})  \label{s1} \\
\text{ }u &=&Y_{1}p  \label{s2}
\end{eqnarray}%
where $Y_{1}$ is the admittance of the fundamental frequency. The
characteristic equation is written as:%
\begin{equation}
F^{\prime }(p_{s})=Y_{1}  \label{s3}
\end{equation}%
As it is well known (see Ref. \cite{grand}), this gives the condition $\
Im(Y_{1})=0$, thus the playing frequency $f_{p}$ at the threshold can be
deduced. Moreover, if at this frequency, $\beta _{1}\overset{def}{=}%
Y_{1}/\zeta $, the pressure threshold is given by:
\begin{equation}
f^{\prime }(\Delta p_{s})=-\beta _{1}  \label{s4}
\end{equation}%
because $F(p)=\zeta f(\Delta p).$ Therefore $\Delta p_{s}$ satisfies Eq. (%
\ref{e914}):
\begin{equation}
\Delta p_{s}=\Delta _{M}\text{ \ and }\gamma _{th}=\Delta _{M}+p_{s}
\label{s5}
\end{equation}%
as expected in section \ref{is}. As a consequence, when the losses depend on
frequency, the value of the threshold is the same as for the Raman model,
and the limit of existence of the two-state regime $\beta _{1}=1$ is
unchanged. Nevertheless the hypothesis has been done that the small
oscillations are sinusoidal (on this subject, see Refs. \cite{worman}, \cite%
{grand} or \cite{ricaud}), and this is not true for inverse bifurcation.
When several harmonics interact, the problem becomes much more intricate,
especially because of resonance inharmonicity. Moreover taking into account
the frequency dependence of the losses leads to a distinction between the
threshold of the fundamental regime and the \textquotedblleft overblown"
regimes (see Ref. \cite{karkar}): this distinction does not exist with the
Raman model, which allows a distinction based upon the initial conditions
only.

As a summary, it can be said that if we suppose that the impedance peak of
the operating frequency is higher than the other ones, the emergence
bifurcation is direct and the limit $\beta _{1}=Y_{1}/\zeta =1$ is valid.
This is true in particular for the first register of a clarinet and a part
of the second register. The large increase of radiation losses at higher
frequency due to the open toneholes lattice (see e.g. Ref. \cite{moers})
does not affect the highest peak, therefore the main result of the present
paper can be extrapolated to a large number of notes of a real clarinet.

\subsubsection{Effect of the reed dynamics}

When the reed dynamics is taken into account as that of a 1 dof oscillator,
the following characteristic equation has been obtained by Silva et al \cite%
{silva}:%
\begin{equation}
Y_{1}=\zeta \sqrt{\gamma }\left[ \frac{1}{1+j\theta /Q_{r}-\theta ^{2}}-%
\frac{1-\gamma }{2\gamma }\right]  \label{e80}
\end{equation}%
where $\theta =\omega /\omega _{r}$, $\omega _{r}$ is the reed-resonance
angular frequency, and $Q_{r}$ its quality factor. The threshold pressure
and frequency can be deduced from this equation, and were studied in this
paper; here we are interested in the limit for which the static regime
becomes always stable, i.e. when $\gamma =1.$ If the input impedance of the
resonator is considered around a resonance frequency $\omega _{1}$, it is
possible to write:%
\begin{equation*}
Y=Y_{1}\left[ 1+jQ_{1}\left( \frac{\omega }{\omega _{1}}-\frac{\omega _{1}}{%
\omega }\right) \right]
\end{equation*}%
where $Q_{1}$ is the quality factor of the resonance. For $\gamma =1,$ the
real part of Eq. (\ref{e80}) leads to the following result:%
\begin{equation}
\beta _{1}=\frac{1}{1-\theta ^{2}+Q_{r}^{-2}\frac{\theta ^{2}}{1-\theta ^{2}}%
}.  \label{e81}
\end{equation}%
For a lossless reed, $\beta _{1}=1/(1-\theta ^{2})>1$: the limit of the
losses in the tube allowed for having a sound is increased by the reed
dynamics, which favors the sound production. But the effect of the reed \
losses is to decrease the limit. It can be concluded that reed losses and
resonator losses act in the same sense concerning the range of parameter
allowing sound production (this conclusion is valid for the
direct-bifurcation case).

\subsection{Discussion about musical consequences for the player}

The previous results can be useful in order to understand and teach
important aspects of the sound control by the instrumentalist. Such an
objective knowledge should largely increase the pedagogical efficiency.
Otherwise the approach of the problems remains more subjective and the
explanations can be lengthy and less clear. One of the most useful aspects
is about \textit{pianissimo} playing. The bifurcation diagrams show that the
players have two possibilities: near the emergence and near the extinction.
The first possibility is used for playing \textit{dolce}, with a quasi
monochromatic sound, but the sound is noisy and cannot be sustained for a
long time, due to high value of the airflow $u$ (see Fig. \ref{pfigA}, case I). The
dynamic is not easy to control because of the steepness of the bifurcation
diagram near $\gamma _{th}$. The second possibility, near $\gamma _{e}$,
conducts to a clean pianissimo, with a sound richer in high harmonics. This
can however only be achieved by crossing the curve (\ref{e23}) in Fig.\ref%
{pertes4}, in order to reach the region II where the extinction bifurcation
is direct. \\This property is usually unknown by the players (the ability of
playing such a \textquotedblleft magical" pianissimo is often attributed
exclusively to the \textquotedblleft quality" of the reed). The beginners
reduce the reed opening ($\zeta )$ by \textquotedblleft biting" the reed and
this causes unwanted effects: the pitch rises considerably, due to the decrease of the effect of the reed flow rate (see  \cite{nantes}, and such a
bending stress can cause a plastic (irreversible) deformation of the reed.
The skilled player can reach region II by increasing the damping due to the
lip and use a high blowing pressure near the extinction threshold, playing
in the reverse way (decreasing the mouth pressure for playing louder, see
Fig. \ref{pfigA}, case II). The lip comes very near to the tip of the reed, with a
moderate lip pressure. This effect is probably similar to an increase of $%
\beta _{1}$, resulting in a displacement of the playing parameters on the $%
(\mu ,\zeta )$ plane exactly in the wanted direction, increasing the value
of $\mu $ and decreasing the value of the parameter $\zeta $ proportionally
to $\sqrt{\beta _{1}}$. The decrease in $\zeta $ is probably due to an
increase of the reed stiffness. Decreasing $\zeta $ without
\textquotedblleft biting" too much could also be achieved by modifying the
hydrodynamics of the airflow entering the channel, in order to increase the
\textit{vena contracta}, but to our knowledge no experimental evidence shows
that the player can indeed modify significantly the \textit{vena contracta}.
Conversely, it seems that the parameter $\beta $ cannot be significantly
controlled by the player (in another way than by modifying the length of the
air column). In real life, this parameter of static airflow resistance may
not be determined only by the length and the diameter of the bore but
certainly also by hydrodynamic effects near the channel, due to the
viscosity of the air. This acts in a similar but possibly stronger way than
the static resistance of the bore. Practical tests show that the effects
predicted in the zones III and IV are indeed observed in some pathological
situations, despite the fact that our theoretical model would require many
meters of a tube of small diameter to obtain such high values of $\beta $.

To include the musician mouth in the model is obviously rather complicated,
even if at low frequencies, the effect of the vocal tract is not important.
Therefore the previous analysis requires some conjectures. Besides the
problems of bifurcation, the analysis of the Raman model permits
establishing some facts useful for the musician:

\begin{itemize}
\item The calculation of the mean flow shows that the most economical
blowing pressure is near the beating threshold in Region I (corresponding to
normal playing). This explains that skilled players can sustain the sound
significantly longer than beginners.

\item The transients are much faster if $\zeta $ is large (see Ref. {\cite%
{Kergomard:95}}); weak reeds help doing this, as well as using a moderate
lip pressure. This simplifies the staccato learning.

\item The effects of leaks in the air column (misplacement of a finger,
defective pads) increase the value of $\beta _{1}$, so that regions 0, and
probably III or IV can be possibly visited (see section \ref{FDL}). Almost
any control can be destroyed over the dynamics (or at least rendering the
dynamic control more difficult), despite of the attempts of the clarinettist
to supply more energy to the instrument by opening the embouchure,
increasing $\zeta $.
\end{itemize}

\section{Conclusion}

\bigskip The present paper is focused on limit cycles corresponding to
two-state regimes, and is a complement to the paper \cite{taillard}, which
was focused on transients. Thanks to a formulation focused on the pressure
difference between mouth and mouthpiece, the effect of the nonlinear
function on the production of the two-state regime can be analyzed, and
especially the role of the losses. The map shown in Fig \ref{pertes4} can
certainly be improved by using more complex models, but we think that some
results are robust. When the reed opening at rest is very small or when the
reed stiffness is very large (i.e. when the dimensionless parameter $\zeta $
is very small), losses can be too large and the sound production becomes
impossible. A complement to this conclusion is the following: for $\zeta $
larger than $1/\sqrt{5}$, when losses increase, the emergence bifurcation
becomes inverse before the sound disappears, and the instrument becomes more
difficult to play. For $\zeta $ smaller than this value, when losses
increase, there is a direct passage from the direct emergence bifurcation to
the absence of sound.

\section{Acknowledgements}

This work was supported by the French National Agency ANR within the
SDNS-AIMV and CAGIMA projects. We thank also the high school ARC-Engineering
in Neuch\^{a}tel. Finally we wish to thank Philippe Guillemain and
Christophe Vergez for fruitful discussions.

\section{Appendix: correction to the Ref. \protect\cite{Dalmont:05}}

The instability threshold $\gamma _{ins}$ of the oscillating regime is given
by the condition $1/C=\zeta ^{2}\beta _{2}$ (see Ineqs. (\ref{e301}), it is
the limit of unstable solutions toward period-doubling of the two-state
regime). This leads to the following equation, if $\Sigma $ and $\Pi $ are
defined by Eq. (\ref{AAA}):

\begin{equation}
\frac{4\Pi }{\zeta ^{2}}+(1+3\Pi )^{2}-3\Sigma ^{2}=\frac{4\beta _{1}}{%
1+\beta \beta _{1}}\Sigma (3\Pi -1)  \label{e21}
\end{equation}%
Together with Eq. (\ref{AZR}) leads to a fourth-order equation in $\Sigma $;
from the solution $\Sigma $ the value the threshold of instability $\gamma
_{ins}$ is deduced by using Eq. (\ref{A}). Another method is to start from
certain values of the parameter $\beta _{1}$ and of the solution $\Sigma $,
then to deduce $\Pi $ using Eq. (\ref{AZR}), then $\beta $, which is
solution of a second order equation

In Ref. \cite{Dalmont:05}, Eq. (A23) was correct, but a factor 4 was missing
in Eq. (A24), the correct equation being the present Eq. (\ref{e21}). Then
Eq. (A25) needs to be corrected by introducing a factor 4 on the right-hand
side, and the factor $(2+3\Pi )$ needs to be replaced by $(12\Pi -1)$ in Eq.
(A28) and similarly $(2+3\Pi _{0})$ needs to be replaced by $(12\Pi _{0}-1)$
in Eq. (A30).

Concerning the limit $\gamma _{ins}=\gamma _{b}$, Eqs. (A32) to (A34) of
Ref. \cite{Dalmont:05} are corrected in Section \ref{LIT} of the present
paper. The last equation gives the coefficient $\beta _{1}$ as a series
expansion of the limit $\zeta _{i}$:%
\begin{equation}
\beta _{1}=\frac{\zeta ^{2}}{2}\left[ 1-\zeta ^{2}+\frac{5}{4}\zeta
^{4}-\zeta ^{6}\right]  \label{216}
\end{equation}%
This expression corrects Eq. (A34) of the previous paper (the correction is
small, because the order 6 in $\zeta $ only is concerned), but this
approximation is much less accurate than the present Eq. (\ref{e238}).

\end{document}